\documentclass[preprintnumbers,
amsmath,amssymb,floatfix,10pt,prd,onecolumn,
superscriptaddress,longbibliography,nofootinbib]{revtex4-2}
\usepackage{bm}
\usepackage{amsfonts}
\usepackage{latexsym}
\usepackage{amsmath}
\usepackage{textcomp}
\usepackage{float}
\usepackage{booktabs}
\usepackage{dcolumn}
\usepackage{dsfont}
\usepackage{ragged2e}
\usepackage{epsfig}
\usepackage[dvipsnames]{xcolor}
\usepackage{hyperref}
\hypersetup{           
	colorlinks=true,                
	breaklinks=true,                
	urlcolor= OrangeRed,                
	linkcolor= magenta,                
	bookmarksopen=false,
	filecolor=black,
	citecolor=red,
	linkbordercolor=blue}
\usepackage{graphicx}
\usepackage{overpic}
\usepackage{orcidlink}
\usepackage[T1]{fontenc}

\begin{document}
	\title{Viability of Big Bang Nucleosynthesis in $f(R,L_m)$ Gravity}

	\author{Kajal Phukan \orcidlink{0009-0000-7630-6485}}
	\email[Corresponding author:]{kajalphukan89@gmail.com}
	\affiliation{%
		Department of Physics, Dibrugarh University, Dibrugarh, Assam, India, 786004}

   \author{Rajdeep Mazumdar \orcidlink{0009-0003-7732-875X}}
	\email {rajdeepmazumdar377@gmail.com}
	\affiliation{%
		Department of Physics, Dibrugarh University, Dibrugarh, Assam, India, 786004}

  \author{Kalyan Malakar\orcidlink{0009-0002-5134-1553}}%
\email{kalyanmalakar349@gmail.com}
\affiliation{Department of Physics, Dibrugarh University, Dibrugarh, Assam, India, 786004}
\affiliation{Department of Physics, Silapathar College, Dhemaji, Assam, India, 787059}

	\author{Kalyan Bhuyan\orcidlink{0000-0002-8896-7691}}%
	\email{kalyanbhuyan@dibru.ac.in}
	\affiliation{%
		Department of Physics, Dibrugarh University, Dibrugarh, Assam, India, 786004}%
	\affiliation{Theoretical Physics Division, Centre for Atmospheric Studies, Dibrugarh University, Dibrugarh, Assam, India 786004}

\begin{abstract}
We examine the viability of Big Bang Nucleosynthesis (BBN) constraints in $f(R,L_m)$ gravity, where the gravitational Lagrangian is an arbitrary function of the the Ricci scalar $R$ and matter Lagrangian density $L_m$. We  derive stringent bounds on the underlying model parameters of four different $f(R,L_m)$ gravity models by examining both the primordial abundances of helium-4 ($^4\text{He}$) and the fractional variations in the neutron--proton freeze out temperature. Our results reveal that $f(R,L_m)$ gravity remains viable under the BBN constraints. With the freeze-out condition offering the most dominat constraint on the parameter space, whereas the bounds from $^4\text{He}$ abundance act as a separate abundance level consistency checks on the modified expansion rate. In summary, the presence of these feasible parameter regions demonstrates that $f(R,L_m)$ gravity models can successfully accommodate primordial element abundances without disrupting standard early Universe dynamics.

\end{abstract}

	\maketitle
    \textbf{Keywords:} Big-Bang Nucleosynthesis, early universe, modified gravity, $f(R,L_m)$ gravity.

\section{Introduction}\label{S1}
Current observation evidence suggest that the the universe, which on large scales is both homogeneous and isotropic, is experiencing an accelerating phase of expansion. This late-epoch acceleration is explained by standard cosmology, which is formalised as the $\Lambda$CDM model and uses a cosmological constant as an effective dark energy. However, this model still holds numerous conflicting theoretical issues. For example, although the $\Lambda$CDM model fits a wide range of empirical observations remarkably well, it does not address the underlying physical basis of dark energy. Other major challenges include the cosmological constant problem, which arises from the enormous discrepancy between the theoretically predicted and observationally inferred vacuum energy densities \cite{WF1,WF2}. Closely related are the fine-tuning and cosmic coincidence problems \cite{carroll1992,sahni2000}, which tends to question why the present dark energy density assumes such a remarkably small value and why its energy density becomes dominant precisely at the current stage of cosmic evolution. The presence of such cosmological issues in the current $\Lambda$CDM model motivates physicists to actively seek alternative cosmological theories, which can overcome such issues with a flawless framework and theoretically strong basis. Different approaches adopted by physicists in finding alternatives included, the introduction of dynamical dark energy models with evolving equations of state \cite{X1,X2,X3,X4,X5}, Vacuum Cold Dark Matter (VCDM) models \cite{Z1,Z2}, entropic cosmological models \cite{Y1,Y2}, and the modified theories of gravity \cite{p14,p14new,p16, p17,p18}.\\
In modified theories of gravity, different geometric frameworks have been formulated by generalizing the Einstein-Hilbert action of general relativity (GR) \cite{p14,p14new,p16, p17,p18}. GR was initially extended by replacing the Einstein--Hilbert action with a more general function of the Ricci scalar, leading to the well-known $f(R)$ theory of gravity \cite{PK15}. It is recognised as one of the oldest and most extensively researched modified gravity theories. Such theories have been extensively studied as viable alternatives for explaining the late-time accelerated expansion of the Universe without invoking dark energy \cite{PK16}. The cosmological viability of $f(R)$ models has been investigated in detail \cite{PK17,PK18}, while several models have also been shown to satisfy local gravity and Solar System constraints \cite{PK19,PK20,PK21,PK22}. Their observational signatures, consistency with the equivalence principle, and implications for dark energy have been discussed in Refs.~\cite{PK23,PK24,PK25,PK26,PK27}. Moreover, viable $f(R)$ models capable of unifying the early inflationary epoch with the present cosmic acceleration have been proposed and successfully tested against local observations \cite{PK28,PK29,PK30}. Additional cosmological applications of $f(R)$ gravity can be found in Refs.~\cite{PK31,PK32,PK33}. A simple extension of the $f(R)$ gravity arises by an explicit coupling between matter and geometry, in which the matter Lagrangian density $L_m$ interacts with the curvature through a generic function \cite{PK34}. This framework was subsequently generalized to allow arbitrary matter--geometry couplings \cite{PK35}, and its cosmological and astrophysical consequences have been explored extensively \cite{PK36,PK37,PK38,PK39,PK40}. An even broader generalization, namely the $f(R,L_m)$ gravity theory, was introduced by Harko and Lobo \cite{PK41}, where the gravitational Lagrangian is an arbitrary function of both the Ricci scalar $R$ and the matter Lagrangian density $L_m$. This theory represents one of the most general extensions of gravity formulated within Riemannian geometry. Similar to other matter--geometry coupling models, it predicts non-geodesic motion for test particles through an additional force induced by the explicit curvature--matter coupling and may lead to violations of the equivalence principle, which are tightly constrained by Solar System observations \cite{PK42,PK43}. More recently, the energy conditions and cosmological implications of $f(R,L_m)$ gravity have been investigated in Refs.~\cite{PK44,PK45}.\\
Although modified gravity theories successfully capture late-time cosmic dynamics, their viability also depends on compatibility with well-established early-Universe phenomena. In this domain, Big Bang Nucleosynthesis (BBN) palys the role of an exceptionally sensitive probe~\cite{bbn11,bbn12}. Taking place during the Universe's hot and dense infant phase, within minutes of the Big Bang, BBN governed the nuclear reactions that synthesized primordial light elements, namely deuterium, helium, and lithium~\cite{bbn13}. As the primordial element synthesis is inherently linked to the cosmic expansion rate, even a subtle departures from standard expansion histories drastically alter predicted abundances. BBN thereby serves as a stringent, model independent testbed for evaluating non-standard cosmological frameworks and modified gravity theories~\cite{bbna1,bbna2,bbna3,bbna4,bbna5}. Consequently, observational BBN data have been widely adopted to constrain non-classical cosmological paradigms. For instance, testing Tsallis cosmology against BBN bounds permits only minimal deviations from standard general relativity~\cite{bbn14}. Analogous restrictions have been derived for Barrow and Kaniadakis cosmologies, as well as extended formulations such as $f(T,L_m)$ and $f(Q,T)$ gravity~\cite{bbn54,s1,s2,s3}. Recent literature has further extended BBN constraints to generalized gravity models: Sultan \textit{et al.}~\cite{sultan2025ftbtgbg} explored parameter spaces in extended teleparallel $f(T,B,T_{\mathcal{G}},B_{\mathcal{G}})$ gravity, whereas Jang \textit{et al.}~\cite{jang2025} constrained energy-momentum squared gravity to restrict non-canonical expansion rates. Furthermore, joint MCMC sampling incorporating BBN alongside cosmic chronometers (CC) and baryon acoustic oscillations (BAO) has effectively constrained torsion-scalar gravity~\cite{ftbphi} and extended teleparallel theories~\cite{alomar2025}, yielding tighter parameter bounds that demonstrate cross-epoch consistency. BBN observations together with CMB and BAO measurements, have also been employed to bound potential variations in the gravitational constant~\cite{lamine2025}.\\
Motivated by these considerations, this work provides to best of our knowledge the first systematic evaluation of Big Bang Nucleosynthesis within $f(R,L_m)$ gravity. While $f(R,L_m)$ theory has been rigorously probed in the context of late-time cosmic acceleration \cite{L1,L2} and early-Universe phenomena such as baryogenesis \cite{B1,B2}, its implications for primordial nucleosynthesis remain virtually unexplored. While our empirical analysis relies on established BBN techniques, the primary originality of this study arises from the underlying gravitational framework. By testing $f(R,L_m)$ gravity during the epoch of primordial element formation, we explore a regime of modified gravity, which has remained unexamined in early-Universe cosmology. Rather than introducing a methodological departure from conventional BBN treatments, this work offers a theoretical assessment of how non-standard matter-geometry couplings reshape nucleosynthesis dynamics, thereby yielding novel constraints on the viability of such modified gravity models.\\
The paper is organized as follows: In Sec.~\ref{S2xx}, we briefly discuss the formalism of $f(R,L_m)$ gravity and, for the same, we derive the modified Friedmann equations. In Sec.~\ref{S3xx}, we discuss the basic framework of Big Bang Nucleosynthesis constraints, and subsequently derive the constraints on the free parameters of different $f(R,L_m)$ gravity models in Sec.~\ref{S4xx}. Finally, in Sec.~\ref{S5xx}, we summarize the main results and present concluding remarks.

\section{$f(R,L_m)$ Gravity and Field Equations} \label{S2xx}
In $f(R, \mathcal{L}_m)$ gravity framework the standard general relativity is generalizes  by introducing an arbitrary function of the Ricci scalar $R$ and the matter Lagrangian density $\mathcal{L}_m$, governed by the action as defined as:
\begin{equation}
S = \int d^4x \sqrt{-g} f(R, \mathcal{L}_m)
\end{equation}
Varying this action with respect to the metric tensor $g_{\mu\nu}$ gives the modified gravitational field equations as:
\begin{equation}
f_R R_{\mu\nu} - \nabla_\mu \nabla_\nu f_R + g_{\mu\nu} \square f_R + \frac{1}{2} (f_{\mathcal{L}_m} \mathcal{L}_m - f) g_{\mu\nu} = \frac{1}{2} f_{\mathcal{L}_m} T_{\mu\nu}
\end{equation}
where we define $f_R \equiv \frac{\partial f}{\partial R}$ and $f_{\mathcal{L}_m} \equiv \frac{\partial f}{\partial \mathcal{L}_m}$ respectively. With the energy-momentum tensor following the standard variational definition given as, $T_{\mu\nu} = \frac{-2}{\sqrt{-g}} \frac{\partial(\sqrt{-g}\mathcal{L}_m)}{\partial g^{\mu\nu}}$. One of the defining characteristic of this theory is the non-minimal coupling between curvature and matter. This coupling shows that the covariant divergence of the energy-momentum tensor is non-zero, such that:
\begin{equation}
\nabla^\mu T_{\mu\nu} = 2 \nabla^\mu [\ln(f_{\mathcal{L}_m})] \frac{\partial \mathcal{L}_m}{\partial g^{\mu\nu}}
\end{equation}
To explore the cosmological implications, we assume a homogeneous, isotropic, and spatially flat Friedmann-Lema\^{i}tre-Robertson-Walker (FLRW) metric, such that we have $ds^2 = -dt^2 + a^2(t)(dx^2 + dy^2 + dz^2)$. Which gives the Ricci scalar as, $R = 6(\dot{H} + 2H^2)$. Substituting this geometric background into the general field equations provieds us the modified Friedmann equations for the given framework. These equations helps to illustrate the cosmic expansion in terms of the matter energy density $\rho$ and pressure $p$, and are given as:
\begin{equation}
3H^2 f_R + 3H\dot{f}_R + \frac{1}{2}(f - f_R R - f_{\mathcal{L}_m} \mathcal{L}_m) = \frac{1}{2} f_{\mathcal{L}_m} \rho
\label{oo1}
\end{equation}

\begin{equation}
\dot{H} f_R + 3H^2 f_R - \ddot{f}_R - 3H\dot{f}_R + \frac{1}{2}(f_{\mathcal{L}_m} \mathcal{L}_m - f) = \frac{1}{2} f_{\mathcal{L}_m} p
\end{equation}
where the dot notation represents differentiation with respect to cosmic time $t$. Here, Eq. (\ref{oo1}) can be re-written as:
\begin{equation}
3H^2=\rho+\rho_{de},
\label{fried1}
\end{equation}
where the effective dark energy density can be defined as
\begin{equation}
\rho_{de}
=
\left(\frac{f_{\mathcal{L}_m}}{2f_R}-1\right)\rho
-\frac{3H\dot f_R}{f_R}
-\frac{1}{2f_R}
\left(f-f_RR-f_{\mathcal{L}_m}\mathcal{L}_m\right).
\label{rhode}
\end{equation}
where, throughout the work, we will consider $L_m =\rho$.

\section{Big Bang Nucleosynthesis Constraints} \label{S3xx}
Big Bang Nucleosynthesis (BBN) takes place during the radiation-dominated era of the early Universe \cite{bbn54,70}. 
During this era the Friedmann equations for standard general relativity (GR) can be approximated as:
\begin{equation}
H^{2} \simeq \frac{\rho_{r}}{3M_{p}^{2}} \equiv H_{\rm GR}^{2},
\label{bbn_fried}
\end{equation}
Here, $\rho_{r}$ represents the energy density due to the relativistic particles present in the universe, while $M_{p}=(8\pi G)^{-1/2}=2.4 \times 10^{18}$ GeV denotes the reduced Planck mass. The radiation energy density can be given as:
\begin{equation}
\rho_{r} = \frac{\pi^{2}}{30} g_{*} T^{4},
\label{rho_r}
\end{equation}
Here, $T$ represents the temperature, and $g_*$ is the effective number of degrees of freedom which can be expressed by $g_*=g(T)\simeq 10$ in the BBN era. Using Eqs.~\eqref{bbn_fried} and \eqref{rho_r}, one finds the Hubble parameter in terms of temperature to be:
\begin{equation}
H(T) = \left(\frac{4\pi^{3}g_{*}}{45}\right)^{1/2}\frac{T^{2}}{M_{P}},
\label{H_T}
\end{equation}
Here, $M_{P} = \sqrt{8 \pi }M_{p} = 1.22 \times 10^{19}$ GeV is the Planck mass. Since radiation energy is conserved, we have $a(t) \propto \sqrt{t}$, with $t$ being the cosmic time, leading to the Hubble parameter, $H=1/(2t)$. This leads to the relation between the temperature as:
\begin{equation}
\frac{1}{t} = \left(\frac{16\pi^{3}g_{*}}{45}\right)^{1/2}\frac{T^{2}}{M_{P}},
\end{equation}
or equivalently $T(t)\propto (t/{\rm s})^{-1/2}\,{\rm MeV}$. During BBN, neutrons are produced via proton–neutron conversion processes \cite{70,71}
\begin{align}
\Gamma_{pn}(T) = {} & (n+\nu_{e}\rightarrow p+e^{-}) 
+ (n+e^{+}\rightarrow p+\bar{\nu}_{e}) \nonumber \\
& + (n\rightarrow p+e^{-}+\bar{\nu}_{e}),
\label{pn_rate}
\end{align}
along with the inverse processes $\Gamma_{np}(T)$. The total interaction rate is thus given by
\begin{equation}
\Gamma_{\rm tot}(T)=\Gamma_{pn}(T)+\Gamma_{np}(T).
\end{equation}
Provided all the particles share a common temperature and remain sufficiently relativistic, the Boltzmann distribution may be used in place of the Fermi–Dirac distribution. Additionally, assuming that the mass of electrons is negligible compared to their energies and the energies of neutrinos, the rate of interactions can be expressed as:  \cite{72,73}
\begin{equation}
\Gamma_{\rm tot}(T) = 8\left(12T^{2}+6QT+Q^{2}\right)AT^{3},
\label{tot_rate}
\end{equation}
In this equation, $Q=m_{n}-m_{p}=1.29\times10^{-3}\,{\rm GeV}$ is the difference in masses of the neutron and proton, and $A=1.02\times10^{-11}\,{\rm GeV}^{-4}$. The mass fraction of primordial helium can be written as:
\begin{equation}
Y_{p}=\frac{2\lambda\,x(T_{f})}{1+x(T_{f})},
\end{equation}
Here, $\lambda=\exp[(T_{f}-T_{n})/\tau]$, such that $T_{f} (\simeq 0.67$ MeV$)$ is the temperature at which weak interactions freeze out \cite{73a}, and $T_{n} (\simeq 0.1$ MeV$)$ marks the beginning of nucleosynthesis \cite{73a}. The the equilibrium neutron-to-proton ratio is given by the quantity $x(T_{f})=\exp(-Q/T_{f})$, while $\tau=880.3\pm1.1\,{\rm s}$ stands for the mean neutron lifetime \cite{73a,74}. The factor $\lambda(T_{f})$ takes into consideration the neutron lifetime in the range of $[T_{f},T_{n}]$. In order to calculate the freeze-out temperature, the comparison between the interaction timescale of $\Gamma_{\rm tot}^{-1}$ and the expansion timescale of $H^{-1}$ is made; where thermal equilibrium exists when $H^{-1}\gg\Gamma_{\rm tot}^{-1}$, but decoupling begins when $H^{-1}\ll\Gamma_{\rm tot}$. The freeze out condition is therefore given by:
\begin{equation}
H(T_{f})=\Gamma_{\rm tot}(T_{f}) \simeq c_{q} T_{f}^{5},
\end{equation}
where $c_{q}\equiv 96A \simeq 9.8\times10^{-10}\,{\rm GeV}^{-4}$ \cite{bbn54,72,73}. Using Eq.~\eqref{H_T}, the freeze-out temperature becomes
\begin{equation}
T_{f}=\left(\frac{4\pi^{3}g_{*}}{45\,c_{q}^{2}M_{P}^{2}}\right)^{1/6} \simeq 0.67 MeV.
\label{Tf_standard}
\end{equation}
Within the framework of modified cosmological, the Hubble parameter departs from its standard GR form, which in turn shifts the freeze-out temperature according to $T_{f}\rightarrow T_{f}+\Delta T_{f}$. As a result, the helium mass fraction is correspondingly modified, with the correction given by:
\begin{equation}
\Delta Y_{p}
=Y_{p}\left[\frac{1-Y_{p}}{2\lambda}\ln\left(\frac{2\lambda}{Y_{p}}-1\right)
-\frac{2T_{f}}{\tau}\right]\frac{\Delta T_{f}}{T_{f}},
\label{deltaY}
\end{equation}
Here, we set $\Delta T(T_{n})=0$, given that $T_{n}$ is determined by the binding energy of deuterium \cite{72,73,75}. Observational constraints place the primordial helium abundance gives \cite{76,76a,76b}:
\begin{equation}
Y_{p}=0.2476, \qquad |\Delta Y_{p}|<10^{-4}.
\label{Yp_obs}
\end{equation}
Here, it is important to emphasized that $\vert{}\Delta Y_{p}\vert{}$ refers to the theoretically predicted shift in the primordial helium mass fraction resulting from the modification to the given cosmological model. The bound $\vert{}\Delta Y_{p}\vert{} < 10^{-4}$ therefore represents the allowed size of this correction, and should not be confused with the statistical error associated with the observed value of $Y_{p}$ \cite{76,76a,76b}. This threshold, $\vert{}\Delta Y_{p}\vert{} < 10^{-4}$, has become a standard theoretical benchmark widely adopted in analytical studies constraining modified gravity models through BBN, and is applied to ensure that any departure from the standard BBN prediction stays negligibly small \cite{76,76a,76b}. In modified gravity theories or in the presence of an additional dark energy component, the Friedmann equation can be written as
\begin{equation}
3M_{p}^{2}H^{2}=\rho_{m}+\rho_{r}+\rho_{\rm DE}.
\end{equation}
During the BBN epoch, the matter contribution is negligible as compared to radiation, giving us:
\begin{equation}
H = H_{\rm GR}\left(1+\frac{\rho_{\rm DE}}{\rho_{r}}\right)^{1/2},
\label{H_mod}
\end{equation}
Here, the $H_{\rm GR}$ stands for the expansion rate in standard cosmology. For $\rho_{\rm DE}\ll\rho_{r}$, this reduces to the form:
\begin{equation}
H \simeq H_{\rm GR}\left(1+\frac{1}{2}\frac{\rho_{\rm DE}}{\rho_{r}}\right).
\end{equation}
Such departure gives rise to a corresponding shift in the freeze-out temperature. Applying the relation $H_{\rm GR}\simeq c_{q}T_{f}^{5}$ together with Eq.~\eqref{Tf_standard}, we get:
\begin{equation}
\frac{\Delta T_{f}}{T_{f}} \simeq
\frac{\rho_{\rm DE}}{\rho_{r}}
\frac{H_{\rm GR}}{10\,c_{q}T_{f}^{5}}.
\label{deltaTf}
\end{equation}
We convert the observational constraint on the primordial helium abundance into a bound on the freeze-out temperature, by making use of Eq.~\eqref{deltaY}. Inserting the standard values of the parameters, specifically $Y_{p} = 0.2476$ along with the corresponding standard BBN values of $\lambda$ and $\tau$ \cite{76,76a,76b}, we get:
\begin{equation}
\Delta Y_{p} \simeq 0.213 \frac{\Delta T_{f}}{T_{f}}.
\label{deltaY_num}
\end{equation}
Substituting the constraint $|\Delta Y_{p}| < 10^{-4}$ from Eq.~\eqref{Yp_obs} into this relation gives us:
\begin{equation}
0.213 \left|\frac{\Delta T_{f}}{T_{f}}\right| < 10^{-4}.
\label{Tf_b}
\end{equation}
Finally, giving us the constraint on the fractional change in Freeze-out Temperature as \cite{76,76a,76b}
\begin{equation}
|\frac{\Delta T_{f}}{T_{f}}| < 4.7\times10^{-4},
\label{Tf_bound}
\end{equation}
Hence, any theoretical prediction for the deviation in the expansion rate must conform to this strict mathematical constraint. We may further define a dimensionless parameter
\begin{equation}
Z \equiv \frac{H}{H_{\mathrm{GR}}} = \left(1+\frac{\rho_{\rm DE}}{\rho_{r}}\right)^{1/2},
\label{z}
\end{equation}
which quantifies the ratio between the expansion rate predicted by the modified model and that of standard General Relativity (GR). A numerical best-fit relation for the primordial helium abundance is provided in Ref.~\cite{31}:
\begin{equation}
Y_p = 0.2485 \pm 0.0006 + 0.0016 \left[(\eta_{10} - 6) + 100(Z - 1)\right],
\end{equation}
with $Z$ defined as in Eq.~(\ref{z}), and the baryon-to-photon ratio given by Ref.~\cite{4}:
\begin{equation}
\eta_{10} \equiv 10^{10} \frac{n_B}{n_\gamma} \simeq 6.
\end{equation}
Applying the observational value $Y_p = 0.2449 \pm 0.0040$ \cite{29}, this translates into the following constraint on $Z$:
\begin{equation}
Z = 1.05 \pm 0.10.
\label{a1}
\end{equation}
Hence, using constraints illustrated by Eq. (\ref{Tf_bound}) \& (\ref{a1}), one can constrain the parameters and also investigate the viability of any given cosmological model against the BBN constraints.

\section{BBN Constraints on $f(R,L_m)$ Gravity} \label{S4xx}
Utilizing the general framework and theoretical bounds outlined in Secs.~(\ref{S2xx}) and~(\ref{S3xx}), we now systematically test the observational viability of Big Bang Nucleosynthesis (BBN) across four distinct functional forms of $f(R,L_m)$ gravity as shown below.

\subsection{Linear Rescaled Curvature and Matter Coupling: $f(R, \mathcal{L}_m) = \alpha R + \beta \mathcal{L}_m$}
\begin{figure*}[htb]
\centerline{
\includegraphics[width=1.02\textwidth]{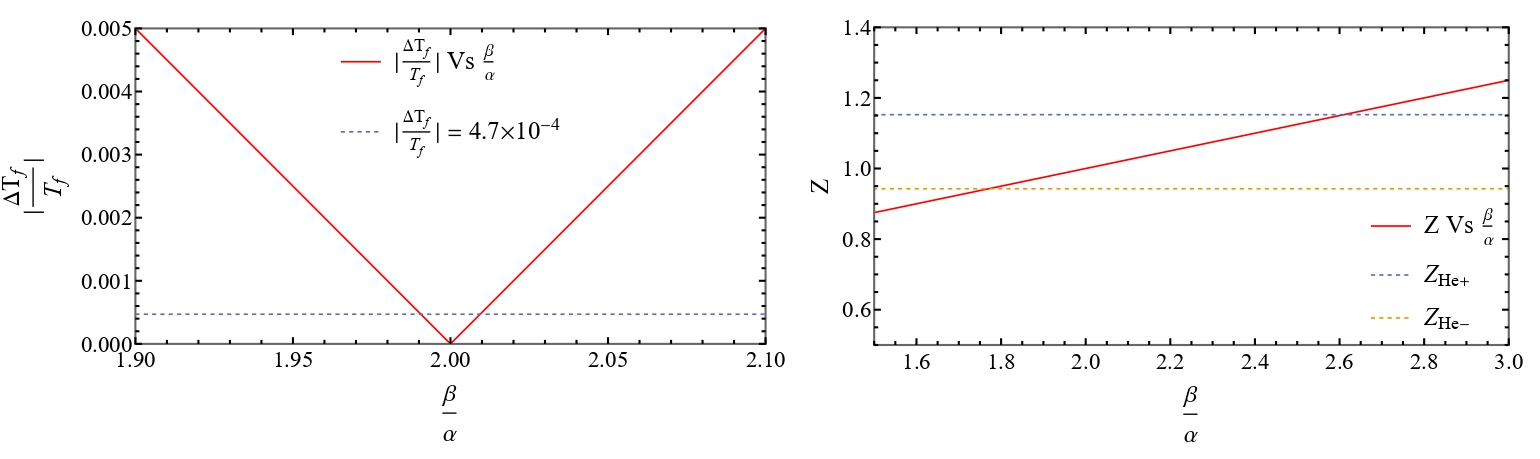}}
\caption{BBN constraints on the parameter $\frac{\beta}{\alpha}$ for Model 1.}
\label{f1}
\end{figure*}

We first study a linear rescale model in which both the Ricci scalar $R$ and the matter Lagrangian density $\mathcal{L}_m$ enter linearly with coupling parameters $\alpha$ and $\beta$ respectively. Taking $\mathcal{L}_m = \rho$, and putting this into Eq.~(\ref{rhode}) gives the effective dark energy density for the given model as:
\begin{equation}
\rho_{\mathrm{DE}} = \frac{1}{2} \left( \frac{\beta}{\alpha} - 2 \right) \rho.
\end{equation}
The universe during the BBN era is radiation dominated, hence taking $\rho \approx \rho_r$ and substituting $\rho_{\mathrm{DE}}$ into Eqs.~(\ref{deltaTf}) and (\ref{z}), we evaluate the fractional change in the freeze-out temperature $\frac{\Delta T_f}{T_f}$ and the expansion-rate parameter $Z \equiv \frac{H}{H_{\mathrm{GR}}}$ as:
\begin{equation}
\frac{\Delta T_f}{T_f} = \frac{\sqrt{g_*} \left(\frac{\beta}{\alpha} - 2\right)}{20 c_q M_p \sqrt{\frac{g_*}{c_q^2 M_p^2}}},
\end{equation}
\begin{equation}
Z = \frac{2\alpha + \beta}{4\alpha} = \frac{1}{2} + \frac{1}{4} \left( \frac{\beta}{\alpha} \right).
\end{equation}
Confronting these analytical expressions against the observational BBN constraints as defined by Eqs. (\ref{Tf_bound}) \& (\ref{a1}), we establish a viable parameter range for $\frac{\beta}{\alpha}$, as illustrated in Fig.~\ref{f1} and summarized in Table~\ref{t1}.
As Fig.~\ref{f1} shows the observational constraints from both weak-interaction freezing out condition and primordial $^4\mathrm{He}$ abundance are mutually compatible with each other, with the freeze-out condition imposing a significantly tighter constraint on the parameter space compared to the primordial $^4\mathrm{He}$ abundance, such that we obtain the final constraint as:
\begin{equation}
1.99 < \frac{\beta}{\alpha} < 2.01
\end{equation}
For the aforementioned gravity model the Standard General Relativity is exactly recovered at $\frac{\beta}{\alpha} = 2$, and the narrow allowed band demonstrates that that given model remains fully BBN viable provided the ratio of the coupling parameters stays within less than $0.5\%$ of its GR value.

\subsection{Non-linear Power-Law Matter Coupling: $f(R, \mathcal{L}_m) = \alpha R + \beta \mathcal{L}_m^n$}
\begin{figure*}[htb]
\centerline{
\includegraphics[width=1.02\textwidth]{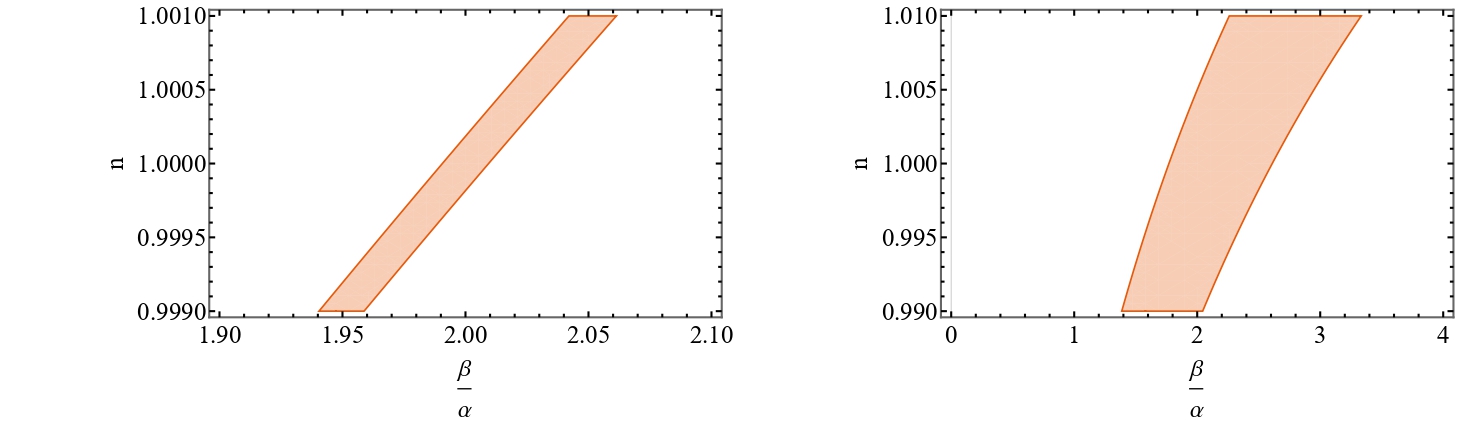}}
\caption{BBN observational constraints on Model 2 in the 2D $(\frac{\beta}{\alpha}, n)$ parameter space.}
\label{f21}
\end{figure*}

\begin{figure*}[htb]
\centerline{
\includegraphics[width=1.02\textwidth]{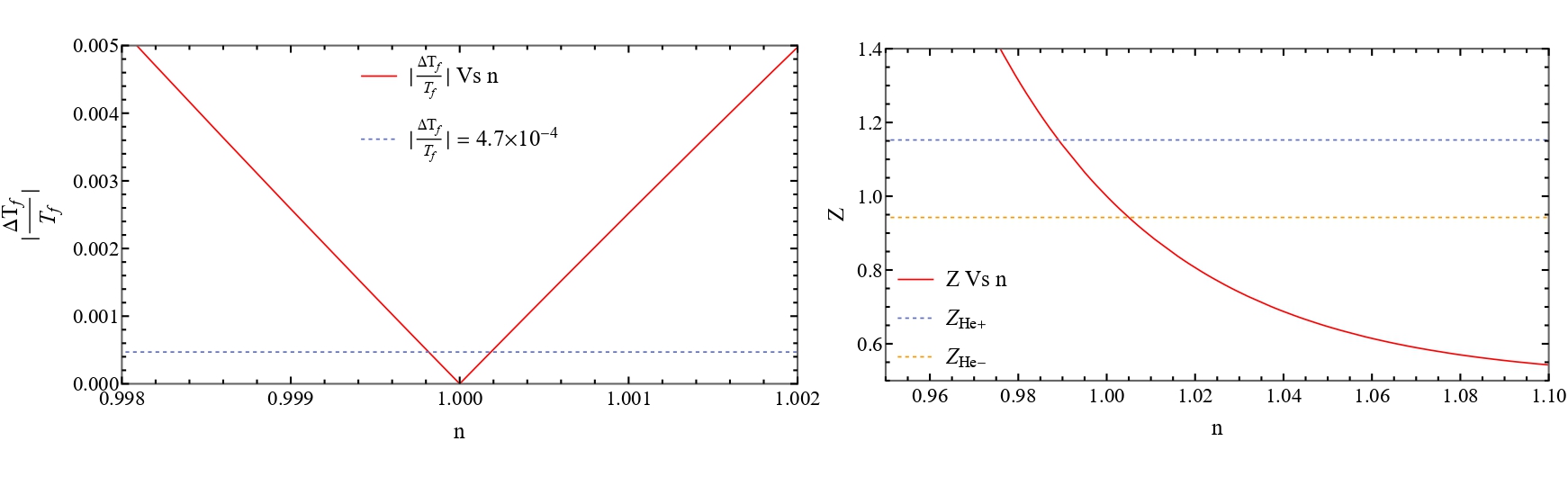}}
\caption{BBN constraints on the non-linear exponent $n$ for Model 2 (with $\frac{\beta}{\alpha} = 2$).}
\label{f2}
\end{figure*}

Next, we investigate a non-linear power-law extension where the matter Lagrangian density enters with an arbitrary exponent $n$. Now, considering $\mathcal{L}_m = \rho$ and using Eq.~(\ref{rhode}) for the given model we obtain the effective dark energy density as:
\begin{equation}
\rho_{\mathrm{DE}} = -\rho + \frac{(2n - 1) \beta}{2\alpha} \rho^n.
\end{equation}
Similar to the first case employing $\rho_{\mathrm{DE}}$ in Eqs.~(\ref{deltaTf}) and (\ref{z}) for the BBN era by taking $\rho \approx \rho_r$, we evaluate $\frac{\Delta T_f}{T_f}$ and the parameter $Z$ as:
\begin{equation}
\frac{\Delta T_f}{T_f} = \frac{3^{\frac{4}{3}-n} 10^{-n-\frac{1}{3}} \left[ \beta 2^{-\frac{2n}{3}} 3^{1-\frac{4n}{3}} 5^{1-\frac{2n}{3}} (2n-1) \pi^{\frac{10n}{3}} \left(g_* \left(\frac{g_*}{c_q^2 M_p^2}\right)^{2/3}\right)^n - \pi^{10/3} \alpha g_* 3^{n-\frac{4}{3}} 10^{n-\frac{2}{3}} \left(\frac{g_*}{c_q^2 M_p^2}\right)^{2/3} \right]}{\pi^{10/3} \alpha \sqrt{g_*} c_q M_p \left(\frac{g_*}{c_q^2 M_p^2}\right)^{7/6}},
\end{equation}
\begin{equation}
Z = \frac{1}{2} \left[ 1 + \frac{\beta}{\alpha} (2n - 1) 2^{-n} 15^{1-n} \pi^{2n-2} \left(g_* T^4\right)^{n-1} \right].
\end{equation}
Confronting these analytical expressions for $\frac{\Delta T_f}{T_f}$ and $Z$ against the observational BBN constraints defined by Eqs.~(\ref{Tf_bound}) and (\ref{a1}), we first try to investigate the presence of any BBN viable parameter space in the $(\frac{\beta}{\alpha}, n)$ plane for Model 2, results for which are as depicted in Fig.~\ref{f21}.
Here, Fig.~\ref{f21} clearly shows that a distinct region exists in the parameter space of the given model, where the model remains fully BBN viable. This allowed parameter space appears significantly tighter under the freeze-out condition compared to the bounds imposed by the $^4\mathrm{He}$ abundance. To further isolate and quantify the constraints on non-linear matter corrections around the General Relativity limit (where $n=1$ and $\frac{\beta}{\alpha}=2$ identically recover GR), we consider the case with $\frac{\beta}{\alpha}=2$, as illustrated in Fig.~\ref{f2} and summarized in Table~\ref{t1}.
As shown in Fig.~\ref{f2}, both observational diagnostics lead to tight constraints on the power-law exponent $n$. However, the freeze-out condition once again provides the more restrictive bound compared to the $^4\mathrm{He}$ abundance, giving us the final constraint as:
\begin{equation}
0.9998 < n < 1.0002
\end{equation}
whereas $^4\mathrm{He}$ abundance permits a broader interval of $0.9891 < n < 1.0050$. This narrow bound confirms that for the given model, non-linear corrections to the matter coupling during the BBN epoch must not deviate by more than $0.018\%$ from the standard linear value ($n=1$) to preserve primordial element synthesis.

\subsection{Quadratic Matter Coupling: $f(R, \mathcal{L}_m) = \alpha R + \mathcal{L}_m + \gamma \mathcal{L}_m^2$}
\begin{figure*}[htb]
\centerline{
\includegraphics[width=1.02\textwidth]{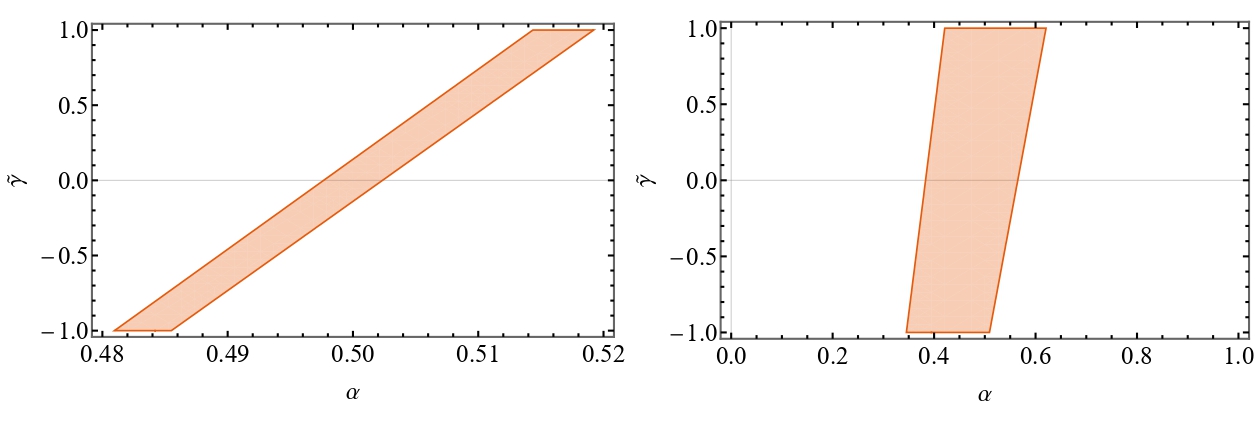}}
\caption{BBN observational constraints on Model 3 in the 2D $(\alpha, \tilde{\gamma})$ parameter space.}
\label{f31}
\end{figure*}

\begin{figure*}[htb]
\centerline{
\includegraphics[width=1.02\textwidth]{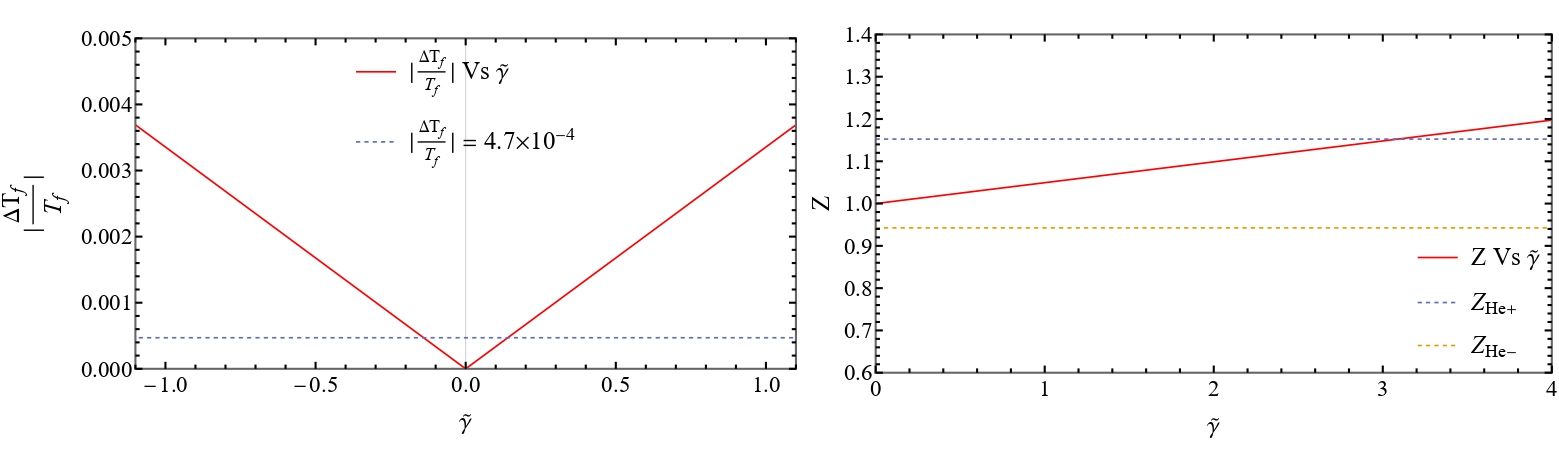}}
\caption{BBN constraints on the rescaled quadratic parameter $\tilde{\gamma}$ (where $\gamma = \tilde{\gamma} \times 10^{10}$) for Model 3 (with $\alpha = 1/2$).}
\label{f5}
\end{figure*}

We now consider a quadratic matter-coupling model in which the leading non-linear correction is quadratic in the matter Lagrangian density, with a coupling constant $\gamma$. Employing Eq.~(\ref{rhode}) and taking $\mathcal{L}_m = \rho$, for the given model we obtain the effective dark energy density as:
\begin{equation}
\rho_{\mathrm{DE}} = \frac{\rho \left( 1 - 2\alpha + 3\gamma \rho \right)}{2\alpha}.
\end{equation}
And, using $\rho_{\mathrm{DE}}$ in Eqs.~(\ref{deltaTf}) and (\ref{z}) with the consideration $\rho \approx \rho_r$ during the BBN era, we obtain:
\begin{equation}
\frac{\Delta T_f}{T_f} = \frac{\sqrt{g_*} \left[ 10 - 20\alpha + \frac{\pi^{10/3} \gamma g_* \left(\frac{g_*}{c_q^2 M_p^2}\right)^{2/3}}{3 \sqrt[3]{3} \, 10^{2/3}} \right]}{200 \alpha c_q M_p \sqrt{\frac{g_*}{c_q^2 M_p^2}}},
\end{equation}
\begin{equation}
Z = \frac{10 + 20\alpha + \pi^2 T^4 \gamma g_*}{40\alpha}.
\end{equation}
Now, introducing a rescaled coupling parameter $\tilde{\gamma}$ defined as $\gamma = \tilde{\gamma} \times 10^{10}$ for numerical convenience and confronting the analytical expressions for $\frac{\Delta T_f}{T_f}$ and $Z$ with the observational BBN constraints as given by Eqs.~(\ref{Tf_bound}) and (\ref{a1}), we first investigate the BBN viable parameter space in the $(\alpha, \tilde{\gamma})$ plane, as depicted in Fig.~\ref{f31}.
Here, Fig.~\ref{f31} clearly demonstrates that a distinct region exists in the $(\alpha, \tilde{\gamma})$ parameter space where the model remains fully BBN viable. With the freeze-out condition imposing the most tighter constraint as compared to the bounds imposed by the primordial $^4\mathrm{He}$ abundance. Subsequently, to isolate the specific effect of the coupling constant of the quadratic term and quantify its bounds around the General Relativity limit, we consider the GR-normalized case with $\alpha = 1/2$, as illustrated in Fig.~\ref{f5} and summarized in Table~\ref{t1}. Under this choice, standard General Relativity is recovered at $\tilde{\gamma} = 0$ (i.e., $\gamma = 0$).
As shown in Fig.~\ref{f5}, both the constraining probes provides compatible parameter bounds, with the freeze-out condition serving as the dominant one. Specifically, the freeze-out condition constraints the coupling parameter to $-1.4 \times 10^9 < \gamma < 1.4 \times 10^9$ (corresponding to $-0.14 < \tilde{\gamma} < 0.14$). In comparison, the $^4\mathrm{He}$ abundance provides a more wider bound of $0 < \gamma < 3.09 \times 10^{10}$ (corresponding to $0 < \tilde{\gamma} < 3.09$). Taking the joint physical intersection of both the constraints the final BBN allowed parameter range is established as:
\begin{equation}
0 < \tilde{\gamma} < 0.14
\end{equation}
which translates to the unscaled interval $0 < \gamma < 1.4007 \times 10^9$. This constraint demonstrates that, for the choice $\alpha=1/2$, the quadratic matter-coupling model remains consistent with BBN constraints provided the coupling parameter remains within the above allowed range.

\subsection{Exponential Matter Coupling: $f(R, \mathcal{L}_m) = \alpha R + \mathcal{L}_m e^{\beta \mathcal{L}_m}$}
\begin{figure*}[htb]
\centerline{
\includegraphics[width=1.02\textwidth]{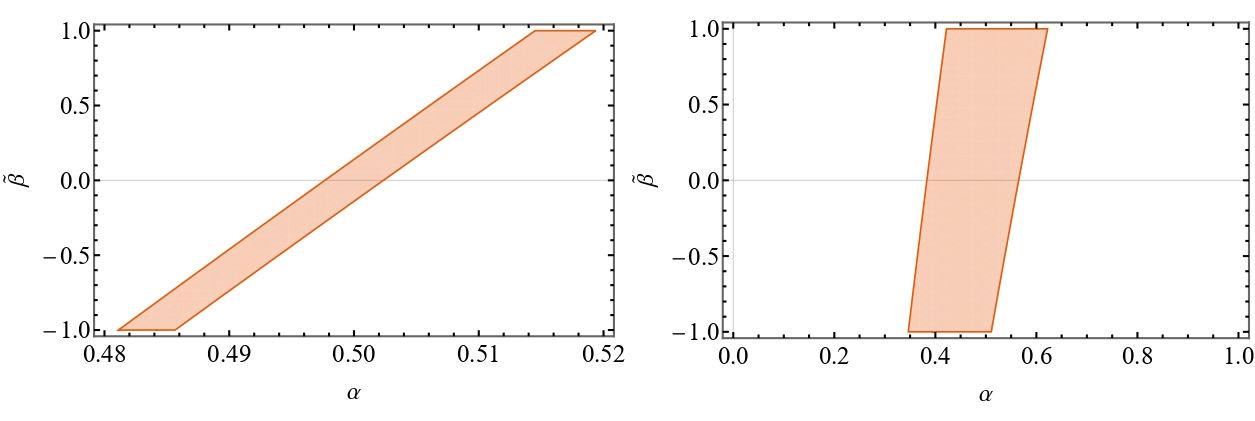}}
\caption{BBN observational constraints on Model 4 in the 2D $(\alpha, \tilde{\beta})$ parameter space.}
\label{f41}
\end{figure*}

\begin{figure*}[htb]
\centerline{
\includegraphics[width=1.02\textwidth]{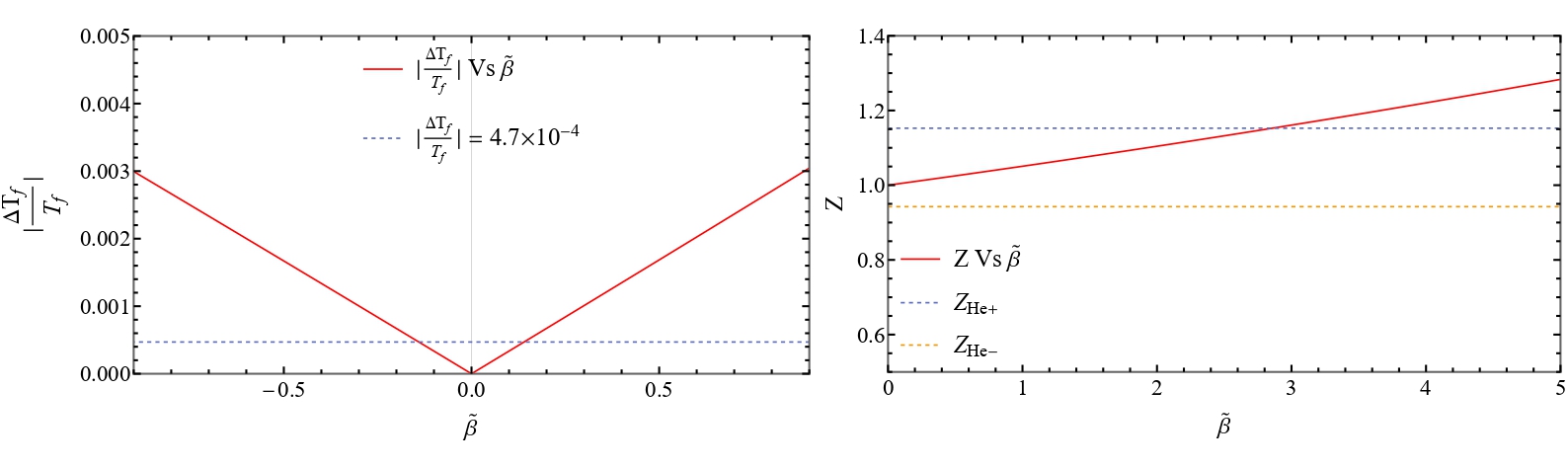}}
\caption{BBN constraints on the rescaled exponential parameter $\tilde{\beta}$ (where $\beta = \tilde{\beta} \times 10^{10}$) for Model 4 (with $\alpha = 1/2$).}
\label{f4}
\end{figure*}

Finally we consider an exponential matter Lagrangian model where non-linear matter corrections enter via an exponential factor modulated by the parameter $\beta$. For which using Eq.~(\ref{rhode}) with $\mathcal{L}_m = \rho$ the effective dark energy density is obtain as:
\begin{equation}
\rho_{\mathrm{DE}} = \frac{1}{2} \rho \left[ -2 + \frac{(1 + 2\beta \rho) e^{\beta \rho}}{\alpha} \right].
\end{equation}
Now, similar to previous model using $\rho_{\mathrm{DE}}$ in Eqs.~(\ref{deltaTf}) and (\ref{z}) with the consideration $\rho \approx \rho_r$ during the BBN era, we get:
\begin{equation}
\frac{\Delta T_f}{T_f} = \frac{\sqrt{g_*} \left[ \frac{1}{\alpha} \left( 1 + \frac{\pi^{10/3} \beta g_* \left(\frac{g_*}{c_q^2 M_p^2}\right)^{2/3}}{45 \sqrt[3]{3} \, 10^{2/3}} \right) \exp\left( \frac{\pi^{10/3} \beta g_* \left(\frac{g_*}{c_q^2 M_p^2}\right)^{2/3}}{90 \sqrt[3]{3} \, 10^{2/3}} \right) - 2 \right]}{20 c_q M_p \sqrt{\frac{g_*}{c_q^2 M_p^2}}},
\end{equation}
\begin{equation}
Z = \frac{1}{4} \left[ 2 + \frac{1}{\alpha} \left( 1 + \frac{1}{15} \pi^2 T^4 \beta g_* \right) \exp\left( \frac{1}{30} \pi^2 T^4 \beta g_* \right) \right].
\end{equation}
Similar to the previous model, we again introduce a rescaled coupling parameter $\tilde{\beta}$ such that $\beta = 10^{10}\tilde{\beta}$, for numerical convenience and confront the analytical expressions for $\Delta T_f/T_f$ and $Z$ with the observational BBN constraints as given by Eqs.~(\ref{Tf_bound}) and (\ref{a1}). Through which we to investigate the presence of any region in the $(\alpha,\tilde{\beta})$ parameter space that satisfies both constraints simultaneously, results for which are shown in Fig.~\ref{f41}.
Here, Fig.~\ref{f41} demonstrates that a viable region exists in the parameter space of the aforementioned model that satisfies the considered BBN constraints and is therefore compatible with the required early-Universe evolution. As found for the previous models, the allowed region is more tightly constrained by the freeze-out condition than by the $^4\mathrm{He}$ abundance, indicating that the former provides the more stringent BBN bound on the model parameters for all cases. Likewise, to isolate the effect of the exponential coupling and quantify its allowed magnitude relative to the General Relativity (GR) limit, we next consider the GR-normalized case with $\alpha=1/2$, as illustrated in Fig.~\ref{f4} and summarized in Table~\ref{t1}. For this choice, the GR limit is recovered when $\tilde{\beta}=0$, equivalently $\beta=0$, allowing the resulting bounds on $\tilde{\beta}$ to be interpreted directly as constraints on deviations from the GR case.
As shown in Fig.~\ref{f4}, the both the constraining probes are overlapping, with the freeze-out condition providing the more stringent bound. Specifically, the freeze-out temperature constraint gives
$-1.402\times10^{9}<\beta<1.399\times10^{9}$, corresponding to
$-0.1402<\tilde{\beta}<0.1399$,
whereas the $^4\mathrm{He}$ abundance allows a broader range,
$0<\beta<2.8565\times10^{10}$, or equivalently
$0<\tilde{\beta}<2.8565$.
The physical intersection of these two constraints therefore give the final BBN-allowed range as:
\begin{equation}
0<\tilde{\beta}<0.1399,
\end{equation}
or, equivalently,
$0<\beta<1.399\times10^{9}$.
Thus, for the GR-normalized choice $\alpha=1/2$, the exponential matter-coupling model remains compatible with the considered BBN constraints provided that the coupling lies within this allowed range.

\begin{table}[htbp]
\centering
\renewcommand{\arraystretch}{1.7}
\setlength{\tabcolsep}{7pt}
\begin{tabular}{>{\centering\arraybackslash}m{2.5cm}
                >{\centering\arraybackslash}m{3 cm}
                >{\centering\arraybackslash}m{3 cm}
                >{\centering\arraybackslash}m{3.5cm}
                >{\centering\arraybackslash}m{3.5cm}}
\hline
\textbf{Constraint} 
& \textbf{Model I} 
& \textbf{Model II} 
& \textbf{Model III} 
& \textbf{Model IV} \\
\hline

\textbf{Freeze-out}
&
$\displaystyle
1.99 < \frac{\beta}{\alpha} < 2.01$
&
$\displaystyle
0.9998 < n < 1.0002$
&
$\displaystyle
-0.14 < \tilde{\gamma} < 0.14$
&
$\displaystyle
-0.1402 < \tilde{\beta} < 0.1399$
\\[4pt]
\hline

\textbf{Primordial $^{4}$He}
&
$\displaystyle
1.77 < \frac{\beta}{\alpha} < 2.61$
&
$\displaystyle
0.9891 < n < 1.0050$
&
$\displaystyle
0 < \tilde{\gamma} < 3.09$
&
$\displaystyle
0 < \tilde{\beta} < 2.8565$
\\[4pt]
\hline

\textbf{Final BBN Constraint}
&
$\displaystyle
1.99 < \frac{\beta}{\alpha} < 2.01$
&
$\displaystyle
0.9998 < n < 1.0002$
&
$\displaystyle
0 < \tilde{\gamma} < 0.14$
&
$\displaystyle
0 < \tilde{\beta} < 0.1399$
\\[4pt]
\hline

\end{tabular}

\caption{Summary of the BBN constraints obtained for the considered
$f(R,\mathcal{L}_m)$ gravity models from the weak-interaction freeze-out
temperature and primordial $^4\mathrm{He}$ abundance. The final BBN-allowed
intervals are obtained from the intersection of the constraints imposed by
the two observational diagnostics. The model parameters have their
respective dimensions to ensure the overall dimensional consistency of the
gravitational action for each functional form of $f(R,\mathcal{L}_m)$.}
\label{t1}
\end{table}


\section{Conclusion}\label{S5xx}
In this study, we investigated the viability of $f(R,L_m)$ gravity within the context of Big Bang Nucleosynthesis (BBN). Our primary objective was to determine whether the modified expansion dynamics induced by various $f(R,L_m)$ formalisms remain consistent with the thermal history required for primordial nucleosynthesis. Because the freeze-out of weak interactions and the resulting abundances of light elements are exceptionally sensitive to variations in the early expansion rate, BBN provides a remarkably stringent test for modified gravitational frameworks. Utilizing the theoretical formulations established in Secs.~(\ref{S2xx}) and (\ref{S3xx}), we derived constraints on four different $f(R,L_m)$ models by analyzing the freeze-out temperature and the primordial $^4\mathrm{He}$ mass fraction.\\
Our findings demonstrate that $f(R,L_m)$ gravity models remain viable under these BBN bounds, exhibiting well-defined regions in parameter space where the predicted $^4\mathrm{He}$ abundance and the fractional shift in the freeze-out temperature align with observational data. Crucially, the freeze-out temperature and primordial abundance analyses offer two complementary probes of modified cosmological dynamics \cite{76c,76d}. While the freeze-out condition directly constrains the Hubble expansion rate at weak-decoupling via $T_f$, the primordial abundance relations probe the cumulative impact of modified expansion across the entire nucleosynthesis epoch via the ratio parameter $Z$ \cite{76c,76d,76e,76f}.\\
Although both diagnostic tools draw upon the same underlying primordial helium observations, they exhibit distinct parametric sensitivities. Because the neutron-to-proton ratio depends exponentially on the freeze-out temperature, even minor deviations in the Hubble expansion rate translate into stringent constraints on $T_f$. Conversely, because the abundance relations rely on phenomenological parameterizations of $Z$, they are relatively less sensitive, yielding broader bounds on the model parameters. Consequently, the freeze-out condition consistently delivers the most stringent BBN constraints across all evaluated models \cite{76c,76d,76e,76f}. In summary, this analysis establishes that $f(R,L_m)$ cosmological models are fully viable during the early Universe, provided their model parameters remain within the bounds dictated by freeze-out and helium observations. The BBN-allowed parameter bounds derived here establish vital early-Universe priors for $f(R, L_m)$ gravity. Key prospective extensions include embedding these modified expansion histories into numerical BBN codes (\texttt{AlterBBN}~\cite{l1}, \texttt{PArthENoPE}~\cite{l2}) for full MCMC likelihood analyses, as well as, combining these primordial bounds with linear perturbation theory and late-time observations (DESI, Pantheon+, CMB) to determine if $f(R, L_m)$ gravity can globally resolve cosmic acceleration and $H_0/S_8$ tensions.
\section*{Data Availability}
All data used in this study are cited
in the references and were obtained from publicly available sources.
\section*{Funding Declaration}
The author(s) received no specific funding for this work.


\end{document}